\documentclass[prd,twocolumn,amsfonts,amssymb]{revtex4}

\usepackage{graphicx}

\usepackage{bm}

\newcommand{\beq}{\begin{equation}}
\newcommand{\eeq}{\end{equation}}
\newcommand{\beqs}{\begin{eqnarray}}
\newcommand{\eeqs}{\end{eqnarray}}

\begin{document}

\title{Scheme-Independent Calculation of $\gamma_{\bar\psi\psi,IR}$ for an 
SU(3) Gauge Theory} 

\author{Thomas A. Ryttov$^a$ and Robert Shrock$^b$}

\affiliation{(a) \ CP$^3$-Origins and Danish Institute for Advanced Study \\
University of Southern Denmark, Campusvej 55, Odense, Denmark}

\affiliation{(b) \ C. N. Yang Institute for Theoretical Physics \\
Stony Brook University, Stony Brook, NY 11794, USA }

\begin{abstract}

We present a scheme-independent calculation of the infrared value of 
the anomalous dimension of the fermion bilinear, 
$\gamma_{\bar\psi\psi,IR}$ in an SU(3) gauge theory as a function of 
the number of fermions, $N_f$, via a series expansion in powers of 
$\Delta_f$, where $\Delta_f=(16.5-N_f)$, to order $\Delta_f^4$.
We perform an extrapolation to obtain the first determination of the 
exact $\gamma_{\bar\psi\psi,IR}$ from continuum field theory.
The results are compared with calculations of the $n$-loop values of 
this anomalous dimension from series in powers of the coupling 
and from lattice measurements. 

\end{abstract}


\maketitle

A fundamental problem in quantum field theory concerns the evolution of an
asymptotically free gauge theory from large Euclidean momentum scales $\mu$ in
the ultraviolet (UV), where it is weakly coupled, to small $\mu$ in the
infrared (IR). The dependence of the running gauge coupling $g=g(\mu)$ on $\mu$
is determined by the beta function \cite{rg}, $\beta = d\alpha/dt$, where
$\alpha(\mu) = g(\mu)^2/(4\pi)$ and $dt=d\ln \mu$ (we often suppress the
argument $\mu$ in the notation).  Here we consider an asymptotically free (AF)
vectorial gauge theory with gauge group $G={\rm SU}(3)$ and $N_f$ fermions
$\psi_i$, $i=1,...,N_f$ in the fundamental ($F$) representation. 
The fermions are
taken to be massless, since a fermion with mass $m$ is integrated out of the
effective theory for $\mu < m$ and hence does not affect the evolution to the
IR with $\mu < m$.  This theory corresponds to quantum chromodynamics (QCD)
with $N_f$ massless quarks.

The beta function of this theory has the series expansion
\beq
\beta = -2\alpha \sum_{\ell=1}^\infty b_\ell \, a^\ell =
-2\alpha \sum_{\ell=1}^\infty \bar b_\ell \, \alpha^\ell \ ,
\label{beta}
\eeq
where $a=g^2/(16\pi^2)=\alpha/(4\pi)$, $b_\ell$ is the $\ell$-loop coefficient,
$\bar b_\ell = b_\ell/(4\pi)^\ell$, and we extract an overall minus sign in
Eq. (\ref{beta}).  The $n$-loop ($n\ell$) beta function, denoted
$\beta_{n\ell}$, is given by Eq. (\ref{beta}) with the upper limit on the
$\ell$-loop summation changed from $\ell=\infty$ to $\ell=n$. The one-loop and
two-loop coefficients are independent of the scheme used for regularization and
renormalization (i.e., scheme-independent, SI), while the $b_\ell$ with 
$\ell \ge 3$ are scheme-dependent (SD) \cite{gross75}; these are 
$b_1=11-(2/3)N_f$
\cite{b1} and $b_2=102-(38/3)N_f$ \cite{b2}.  In our analysis, we formally
extend $N_f$ to nonnegative real numbers, understanding that the physical
values are nonnegative integers. Since $b_1$ vanishes as $N_f$ increases
through the value $N_{f,b1z}=33/2$, the AF property implies the upper bound
$N_f < N_{f,b1z}=33/2$, which we assume.  The interval $0 \le N_f < N_{f,b1z}$
is denoted $I_{AF}$. We define
\beq
\Delta_f = N_{f,b1z}-N_f = \frac{33}{2}-N_f \ . 
\label{Deltaf}
\eeq
The coefficients $b_3$ and $b_4$ were calculated in \cite{b3}
and \cite{b4} (and checked in \cite{b4p}), 
in the $\overline{\rm MS}$ scheme \cite{msbar}; e.g., 
$b_3 = (2857/2)-(5033/18)N_f+(325/54)N_f^2$. 

As $N_f \in I_{AF}$ increases from 0, $b_2$ decreases, vanishing at
$N_{f,b2z}=153/19=8.0526$, and is negative in the interval $153/19 < N_f <
33/2$, which is denoted $I_{IRZ}$.  If $N_f \in I_{IRZ}$, then the two-loop
beta function $\beta_{2\ell}$ has an IR zero (IRZ), at
$\alpha=\alpha_{IR,2\ell}=-4\pi b_1/b_2$.  Here the IR zero of the $n$-loop
beta function $\beta_{n\ell}$ is denoted $\alpha_{IR,n\ell}$.  As $N_f \nearrow
N_{f,b1z}$ at the upper end of $I_{IRZ}$, $\alpha_{IR,2\ell} \to 0$, enabling a
perturbative study of the IR behavior \cite{b2,bz}. As $N_f \in I_{IRZ}$
decreases below $N_{f,b1z}$, $\alpha_{IR,2\ell}$ increases, eventually becoming
O(1). Therefore, the perturbative study of IR behavior for $N_f$ toward the
middle and lower part of $I_{IRZ}$, necessitates higher-loop
calculations. These were performed to four-loop order in
\cite{gk}-\cite{irgen}.  For $n \ge 3$ loops, $\alpha_{IR,n\ell}$ is
scheme-dependent, and the effect of this was studied in \cite{sch}.  For
sufficiently large $N_f \in I_{IRZ}$, the theory evolves to an exact IR fixed
point (IRFP) of the renormalization group (RG) in a chirally symmetric
non-Abelian Coulomb phase (NACP).  For sufficiently
small $N_f$, spontaneous chiral symmetry breaking (S$\chi$SB) occurs, the
fermions gain dynamical masses, and they are integrated out of the low-energy
effective theory that is applicable at lower scales in the IR. In this latter
case, the IR zero is only an approximate IRFP.  The lowest value of $N_f$ in
the NACP is denoted as $N_{f,cr}$. The UV to IR flow in the
chirally broken phase near to this lower boundary of the NACP can exhibit
quasiconformal behavior, which might be relevant to physics beyond the Standard
Model. It is of great interest to elucidate the properties of the
theory at the IRFP. 

In this letter we report a significant advance toward the achievement of this
goal, namely a new scheme-independent calculation of the anomalous dimension of
the fermion bilinear, $\bar\psi\psi \equiv \bar\psi_i\psi_i$ (no sum on $i$),
evaluated at the IR zero of the beta function. We denote this as
$\gamma_{\bar\psi\psi,IR}$ \cite{gammam}. As a physical quantity, this is
clearly scheme-independent \cite{gross75}.  The
full scaling dimension of the $\bar\psi\psi$ operator is
$D(\bar\psi_i\psi_i)=3-\gamma_{\bar\psi\psi}$, with the anomalous dimension
$\gamma_{\bar\psi\psi}=-d\ln Z_{\bar\psi\psi}/dt$, where $Z_{\bar\psi\psi}$ is
the renormalization constant for this operator.  For brevity, we set
$\gamma_{\bar\psi\psi} \equiv \gamma$ and 
$\gamma_{\bar\psi\psi,IR} \equiv \gamma_{IR}$.
In a usual perturbative
calculation, $\gamma$ is expressed as the series
\beq
\gamma = \sum_{\ell=1}^\infty c_\ell a^\ell = 
                        \sum_{\ell=1}^\infty \bar c_\ell \alpha^\ell \ , 
\label{gamma_series}
\eeq
where $c_\ell$ is the $\ell$-loop term and $\bar c_\ell = c_\ell/(4\pi)^\ell$.
The coefficient $c_1=8$ is scheme-independent, while the $c_\ell$ with $\ell
\ge 2$ are scheme-dependent and have been calculated to $\ell=4$ loop order in
\cite{c4}.  The $n$-loop result for $\gamma$ is defined by replacing
$\ell=\infty$ by $\ell=n$ as the upper limit on the sum in
(\ref{gamma_series}), and the $n$-loop approximation to the exact
$\gamma_{IR}$, denoted $\gamma_{IR,n\ell}$, is then obtained by setting 
$\alpha = \alpha_{IR,n\ell}$ in $\gamma_{n\ell}$. A rigorous upper bound is
\beq
\gamma_{IR} < 2
\label{gamma_upperbound}
\eeq
in both the NACP and the chirally broken phase \cite{gammabound}. 

The quantities $\alpha_{IR,n\ell}$ and $\gamma_{IR,n\ell}$ were calculated to
$n=4$ loop order in \cite{bvh,ps}. Although $b_5$ and $c_5$ have not yet been
calculated for general $G$ and fermion representation $R$, $c_5$ is known
\cite{c5su3} and $b_5$ has recently been calculated \cite{b5su3} in the
$\overline{\rm MS}$ scheme for the present theory, $G={\rm SU}(3)$, $R=F$.
Using these results, we have computed $\alpha_{IR,5\ell}$ and
$\gamma_{IR,5\ell}$ in this scheme \cite{flir}.

It is highly desirable to construct a calculational framework in which
$\gamma_{IR}$ can be expressed as a series expansion such that at
every order in this expansion, the result is scheme-independent. One of us
(T.A.R.) recently achieved this goal in \cite{gtr}, expressing 
$\gamma_{IR}$ as
\beq
\gamma_{IR} = \sum_{k=1}^\infty \kappa_k \Delta_f^k  \ ,
\label{gamma_delta_series}
\eeq
where each $\kappa_k$ is scheme-independent. 
The inputs for the calculation of $\kappa_k$ are the $b_\ell$ at loop order 
$1 \le \ell \le k+1$ and the $c_\ell$ at loop order
$1 \le \ell \le k$.  For the finite series approximation we denote 
$\gamma_{IR,\Delta^p} = \sum_{k=1}^p \kappa_k \Delta_f^k$. 
Ref. \cite{gtr} gave $\gamma_{IR,\Delta^p}$ for the powers $1 \le p \le 3$
for general $G$ and $R$. 

Here we report two new results: (i) the calculation of 
$\kappa_4$ and hence $\gamma_{IR,\Delta^4}$, and (ii) using
the $\gamma_{IR,\Delta^p}$ with $p$ up to 4, an 
extrapolation to the exact $\gamma_{IR}$
for $G={\rm SU}(3)$, $R=F$, and $N_f \in I_{IRZ}$. The lower-order
coefficients for this SU(3) theory are \cite{kappafactors}
\beq
\kappa_1 = \frac{16}{3 \cdot 107} = 4.9844 \times 10^{-2} 
\label{kappa1}
\eeq
\beq
\kappa_2 = \frac{125452}{(3 \cdot 107)^3} = 3.7928 \times 10^{-3} 
\label{kappa2}
\eeq
and
\beq
\kappa_3 = \frac{972349306}{(3 \cdot 107)^5} - 
\frac{140800}{3^3 \cdot (107)^4} \, \zeta(3) = 2.3747 \times 10^{-4}
\label{kappa3}
\eeq
Using the SI method of \cite{gtr} together with $b_5$ from \cite{b5su3} 
(and lower-loop $b_\ell$ and $c_\ell$), we find 
\beqs
\kappa_4 & = & \frac{33906710751871}{2^2 (3 \cdot 107)^7} 
-\frac{1684980608}{3^5 \cdot (107)^6} \, \zeta(3)
+ \frac{59840000}{(3 \cdot 107)^5} \, \zeta(5) \cr\cr
& = & 3.6789 \times 10^{-5} \ , 
\label{kappa4}
\eeqs
where $\zeta(s)= \sum_{n=1}^\infty n^{-s}$ is the Riemann zeta function.

\begin{figure}
  \begin{center}
    \includegraphics[height=6cm]{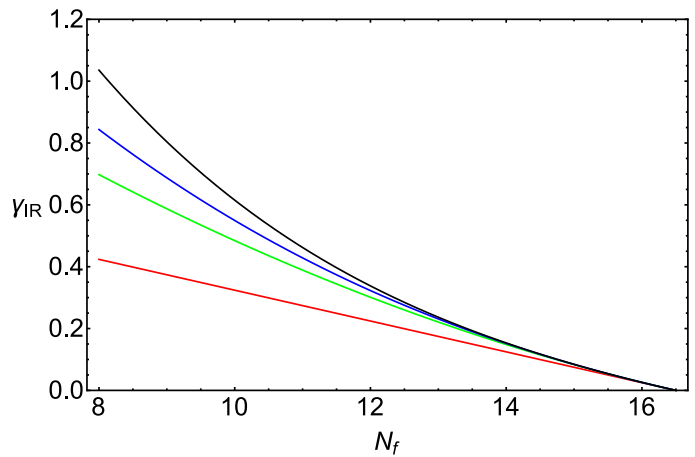}
  \end{center}
\caption{Plot of $\gamma_{IR,\Delta^p}$ for $1 \le p \le 4$ as a
  function of $N_f$. From bottom to top, the curves (with colors online) 
  refer to $\gamma_{IR,\Delta}$ (red), 
           $\gamma_{IR,\Delta^2}$ (green), 
           $\gamma_{IR,\Delta^3}$ (blue), and 
           $\gamma_{IR,\Delta^4}$ (black).}
\label{gamma_Delta_plot}
\end{figure}

In Fig. \ref{gamma_Delta_plot} we show a plot of 
$\gamma_{IR,\Delta^p}$ and in 
Table \ref{gamma_values} we list numerical results 
for $1 \le p \le 4$, with $N_f \in I_{IRZ}$. For comparison, 
this table also lists results for $\gamma_{IR,n\ell}$ at 
$n$-loop level for $1 \le n \le 5$ from \cite{bvh,flir}.  The values 
of $\gamma_{IR,2\ell}$ for $N_f=9, \ 10$ exceed the upper bound
(\ref{gamma_upperbound}) and hence, as noted in \cite{bvh},
we regard these $N_f$ values at the lower
end of $I_{IRZ}$ to be beyond reliable perturbative analysis via the series
(\ref{gamma_series}).  The estimates of $\gamma_{IR,5\ell}$ for $N_f=9, \ 10$
were not given in \cite{flir}; they 
use the IR zero from the $[3,1]$ Pad\'e approximants for $\beta_{5\ell}$. 
Here we see another merit of the SI expansion (\ref{gamma_delta_series}),
namely that it allows us to study the IR behavior closer to the lower end of 
the interval $I_{IRZ}$.  Although 
$N_f=8 < N_{f,b2z}$ is below the lower end of $I_{IRZ}$, we mention that 
$\gamma_{IR,\Delta^p}=0.424, \ 0.698, \ 0.844, \ 1.04$ for $1 \le p \le 4$. 

Having the four SI values $\gamma_{IR,\Delta^p}$ with $1 \le p \le 4$, we can
carry out a polynomial extrapolation to estimate the exact $\gamma_{IR} =
\lim_{p \to \infty} \gamma_{IR,\Delta^p}$ for each $N_f$.  We have investigated
two such extrapolations, one of which uses all four terms and the other of
which uses the three highest-order terms, i.e. $p=2, \ 3, \ 4$. These two types
of extrapolations give consistent results. 
We report the values obtained with the second extrapolation
method here. For example, for $N_f=12$, we obtain the fitting polynomial
$\gamma_{IR,\Delta^p,fit} = 0.2048p^{-2} -0.3005p^{-1} + 0.400$, 
from which we get 
$\gamma_{IR}=\lim_{p \to \infty} \gamma_{IR,\Delta^p,fit}=0.400$ for
this $N_f$. We list our results for $\gamma_{IR}$ as a function
of $N_f$ from this extrapolation in Table \ref{gamma_values}. For $N_f$ values
near the upper end of the interval $I_{IRZ}$, where $\Delta_f$ is small, our
$\gamma_{IR,\Delta^4}$ and extrapolation to the exact $\gamma_{IR}$
(both of which are SI) are very close to the value of
$\gamma_{IR,4\ell}$ calculated in the $\overline{\rm MS}$ scheme
\cite{bvh,ps} and in other schemes \cite{sch} and to the value of 
$\gamma_{IR,5\ell}$ in \cite{flir}.  As $N_f$ decreases in $I_{IRZ}$ ,
our $\gamma_{IR,\Delta^5}$ and extrapolated exact $\gamma_{IR}$ become 
progressively  larger than the corresponding values of $\gamma_{IR,n\ell}$ for 
$3 \le n \le 5$.  If we were to apply the same extrapolation procedure at 
$N_f=8$ (below $N_{f,b2z}$), we would get an unphysical value of $\gamma_{IR}$
slightly above 2. 

An important general result concerns the monotonicity of $\gamma_{IR}$ as a
function of $N_f$.  We find that for $G={\rm SU}(N_c)$ for general $N_c$ and
for $R$ equal to the fundamental, adjoint, and symmetric and antisymmetric
rank-2 tensor representations, the $\kappa_p$ for $p=1, \ 2, \ 3$ given in 
\cite{gtr} are positive.  Hence, for all of these cases, for $p=1,2,3$, 
$\gamma_{IR,\Delta^p}$ is a monotonically increasing function of 
$\Delta_f$, i.e., a monotonically decreasing function of $N_f$ in 
the range where this $\Delta_f$ expansion applies, which includes the
interval $I_{IRZ}$. Further, our Eq. (\ref{kappa4}) shows that 
$\kappa_4 > 0$ $G={\rm SU}(3)$ and $R=F$, so for this case 
$\gamma_{IR,\Delta^4}$ and our extrapolated exact $\gamma_{IR}$ are also 
monotonically increasing functions of $\Delta_f$, i.e. decreasing functions of
$N_f$, throughout $I_{IRZ}$.  A plausible conjecture, based on these
results, is that for $G={\rm SU}(N_c)$ with general $N_c$ and for 
$R=F$, $\kappa_p > 0$ for all $p \ge 1$. Assuming this
conjecture is correct, then the inequality 
$\gamma_{IR,\Delta^p} \le \gamma_{IR}$ follows (realized as a strict inequality
except at $N_f = N_{f,b1z}$ where $\gamma_{IR} = 0$).  
We note that $\kappa_p > 0$ for all $p \ge 1$ in QCD with
${\cal N}=1$ supersymmetry (SQCD) \cite{nsvz,gtr}.

We next compare our results for $\gamma_{IR,\Delta^4}$ and extrapolation for
$\gamma_{IR}$ with lattice measurements of $\gamma_{IR}$ \cite{lgtreviews}.
The most extensive measurements have been carried out for $N_f=12$ and range
from $\gamma_{IR} \sim 0.4$ to $\gamma_{IR} \simeq 0.2$ \cite{lsd}-\cite{kuti}.
Our value $\gamma_{IR,\Delta^4}=0.338$ and our extrapolated
$\gamma_{IR} = 0.40$ are consistent with this range of lattice measurements and
are somewhat higher than the five-loop value $\gamma_{IR,5\ell}=0.255$ from the
conventional $\alpha$ series that we obtained in \cite{flir}.  There is also
consistency between our determinations of $\gamma_{IR}$ and rough 
estimates that $\gamma_{IR} \sim 1$ from lattice studies for $N_f=10$ 
\cite{lsdnf10} and $N_f=8$ \cite{latkminf8,lsdnf8}. 

Combining the upper bound $\gamma_{IR} < 2$ with the monotonicity of
$\gamma_{IR}$, we infer that if $\gamma_{IR}$ saturates its upper bound as 
$N_f \searrow N_{f,cr}$ at the lower end of the NACP \cite{gammasqcd}, 
then we would conclude that $8 < N_{f,cr} < 9$.  However, we
stress that it is not known if, in fact, $\gamma_{IR}$ saturates its upper
bound in this way as $N_f \searrow N_{f,cr}$. 

In contrast to $\gamma_{IR}$, the IR zero of $\beta$,
$\alpha_{IR}$, is scheme-dependent.  Nevertheless, one can use the 
$\Delta_f$ expansion to obtain an estimate of $\alpha_{IR}$ that is
complementary to the estimate from the calculation of the zero of $\beta$
expressed as a series expansion in powers of $\alpha$.  We 
write
\beq
\alpha_{IR} = 4\pi \sum_{n=1} \tilde a_n \Delta_f^n \ . 
\label{alpha_delta_series}
\eeq
We have calculated the $\tilde a_n$ for general $G$ and $R$ for 
$1 \le n \le 3$. Using $b_5$ from \cite{b5su3} for $G={\rm SU}(3)$ and 
$R=F$, we have also calculated $\tilde a_4$ for this case, for which we
find 
\beq
\tilde a_1 = \frac{2}{3 \cdot 107} = 0.62305 \times 10^{-2} 
\label{a1tilde}
\eeq
\beq
\tilde a_2 = \frac{11675}{2(3 \cdot 107)^3} = 1.7649 \times 10^{-4}
\label{a2tilde}
\eeq
\medskip
\beqs
\tilde a_3 & = & \frac{145645559}{2^2 \cdot 3^4 \cdot (107)^5} + 
\frac{170720}{3^3 \cdot (107)^4} \, \zeta(3) = 0.90035 \times 10^{-4} \cr\cr
& & 
\label{a3tilde}
\eeqs
\beqs
\tilde a_4 & = & \frac{119816461287557}{2^5 \cdot 3^8 \cdot (107)^7} 
+ \frac{15442747864}{3^7 \cdot (107)^6} \, \zeta(3) 
- \frac{24534400}{(3 \cdot 107)^5} \, \zeta(5) \cr\cr
& = & 1.7453 \times 10^{-6} \ . 
\label{a4tilde}
\eeqs

In summary, using the recently calculated $b_5$ from \cite{b5su3}, we have 
presented a scheme-independent calculation of $\gamma_{IR,\Delta^4}$ and an
extrapolation to estimate the exact anomalous dimension of the fermion bilinear
at the IR zero of the beta function, $\gamma_{IR}$, as a function of $N_f$ in 
a QCD-like gauge theory. We have compared the results with $n$-loop
calculations obtained from power series in the coupling and with lattice
measurements. 

The research of T.A.R. and R.S. was supported in part by the Danish National
Research Foundation grant DNRF90 to CP$^3$-Origins at SDU and by the U.S. NSF
Grant NSF-PHY-13-16617, respectively.

\begin{widetext}
\begin{table}
\caption{\footnotesize{Values of the scheme-independent 
IR anomalous dimension for the fermion bilinear,
$\gamma_{IR,\Delta^p}$ for $1 \le p \le 4$ as a function of 
$N_f \in I_{IRZ}$, and the extrapolated values of the exact 
$\gamma_{IR}$, where the number in
parentheses is an estimate of the uncertainty in the last significant figure in
the extrapolated value.  For comparison, we also include 
$\overline{\rm MS}$ calculations of $\gamma_{IR,n\ell}$ at the 
$2 \le n \le 5$ loop level from \cite{bvh,flir}.}}
\begin{center}
\begin{tabular}{|c|c|c|c|c|c|c|c|c|c|} \hline\hline
$N_f$ & $\gamma_{IR,2\ell}$ & $\gamma_{IR,3\ell}$ & $\gamma_{IR,4\ell}$
& $\gamma_{IR,5\ell}$ & $\gamma_{IR,\Delta}$ 
& $\gamma_{IR,\Delta^2}$ & $\gamma_{IR,\Delta^3}$ & $\gamma_{IR,\Delta^4}$ 
& $\gamma_{IR}$ \\ \hline
9  & $> 2$  & 1.062 &  $< 0$  & $< 0$  & 0.374  & 0.587  & 0.687  & 0.804 
& 1.4(2)    \\
10 & $> 2$  & 0.647  & 0.156  & 0.211  & 0.324  & 0.484  & 0.549  & 0.615  
& 0.95(6)   \\
11 & 1.61   & 0.439  & 0.250  & 0.294  & 0.274  & 0.389  & 0.428  & 0.462  
& 0.62(2)   \\
12 & 0.773  & 0.312  & 0.253  & 0.255  & 0.224  & 0.301  & 0.323  & 0.338  
& 0.400(5)  \\
13 & 0.404  & 0.220  & 0.210  & 0.239  & 0.174  & 0.221  & 0.231  & 0.237 
& 0.257(5)  \\
14 & 0.212  & 0.146  & 0.147  & 0.154  & 0.125  & 0.148  & 0.152  & 0.153  
& 0.154(4)  \\
15 & 0.0997 & 0.0826 & 0.0836 & 0.0843 & 0.0748 & 0.0833 & 0.0841 & 0.0843 
& 0.0841(2) \\
16 & 0.0272 & 0.0258 & 0.0259 & 0.0259 & 0.0249 & 0.0259 & 0.0259 & 0.0259 
& 0.0259(1) \\
\hline\hline
\end{tabular}
\end{center}
\label{gamma_values}
\end{table}
\begin{table}
\caption{\footnotesize{Values of $\alpha_{IR,\Delta^p}$ with $1 \le p \le 4$ 
as functions of $N_f \in I_{IRZ}$, together with $\alpha_{IR,2\ell}$ and 
$\overline{\rm MS}$ values of $n$-loop $\alpha_{IR,n\ell}$ 
with $3 \le n \le 5$ for comparison. The values of 
$\alpha_{IR,5\ell}$ for $9 \le N_f \le 12$ are from the $[3,1]$ Pad\'e
approximants (PAs) to the respective beta functions in \cite{flir}.}}
\begin{center}
\begin{tabular}{|c|c|c|c|c|c|c|c|c|} \hline\hline
$N_f$ & $\alpha_{IR,2\ell}$ & $\alpha_{IR,3\ell}$ & $\alpha_{IR,4\ell}$
& $\alpha_{IR,5\ell}$ 
& $\alpha_{IR,\Delta}$ 
& $\alpha_{IR,\Delta^2}$ 
& $\alpha_{IR,\Delta^3}$ 
& $\alpha_{IR,\Delta^4}$ 
\\ \hline
9  & 5.24  & 1.028 & 1.072 & 1.02$_{PA}$ & 0.587 & 0.712 & 1.19  & 1.26  \\
10 & 2.21  & 0.764 & 0.815 & 0.756$_{PA}$ & 0.509 & 0.603 & 0.913 & 0.952 \\
11 & 1.23  & 0.578 & 0.626 & 0.563$_{PA}$ & 0.431 & 0.498 & 0.686 & 0.706 \\
12 & 0.754 & 0.435 & 0.470 & 0.4075$_{PA}$ & 0.352 & 0.397 & 0.500 & 0.509 \\
13 & 0.468 & 0.317 & 0.337 & 0.406  & 0.274 & 0.301 & 0.350 & 0.353 \\
14 & 0.278 & 0.215 & 0.224 & 0.233  & 0.196 & 0.210 & 0.227 & 0.228 \\
15 & 0.143 & 0.123 & 0.126 & 0.127  & 0.117 & 0.122 & 0.126 & 0.126 \\
16 & 0.0416& 0.0397& 0.0398& 0.0398 & 0.0391 & 0.0397 & 0.0398 & 0.0398 \\
\hline\hline
\end{tabular}
\end{center}
\label{alfir_nloop_values}
\end{table}
\end{widetext}


\begin{thebibliography}{99}
\bibitem{gammam}
%
The anomalous dimension $\gamma_{\bar\psi\psi}$ is often denoted $\gamma_m$;
we use the more general notation $\gamma_{\bar\psi\psi}$ here, since no mass
$m$ is generated in the non-Abelian Coulomb phase.
\bibitem{kappafactors}
%
Aside from $\kappa_1$, the numerators of terms in these $\kappa_p$ do not 
have similarly simple factorizations; e.g.,
in $\kappa_2$, $125452=2^2 \cdot 79 \cdot 397$, 
in $\kappa_3$, $972349306 = 2 \cdot 17 \cdot 569 \cdot 50261$, etc.
\bibitem{gammasqcd}
In SQCD, the upper bound on $\gamma_{IR}$ is $\gamma_{IR} < 1$ in the NACP, and
$\gamma_{IR}$ does saturate this upper bound as the the number $N_f$ of 
pairs of chiral superfields $\hat\Phi$ and $\tilde{\hat\Phi}$ decreases toward
the lower end of the NACP.
\end{thebibliography}
\end{document}